\documentclass[prb,twocolumn,floatfix,showpacs]{revtex4}

\pdfoutput=1

\usepackage{amsmath,amsfonts,amssymb,bm}
\usepackage{dcolumn}
\usepackage[final]{graphicx}
\usepackage{bm}

\begin{document}

\bibliographystyle{apsrev}

\title{Shortcoming of spectral filtering in semiconductor-based entangled photon sources}

\author{Gernot Pfanner}
\author{Marek Seliger}
\author{Ulrich Hohenester}%\email{ulrich.hohenester@uni-graz.at}
\affiliation{Institut f\"ur Physik,
  Karl--Franzens--Universit\"at Graz, Universit\"atsplatz 5,
  8010 Graz, Austria}

\date{\today}

\begin{abstract}

We theoretically investigate the production of polarization-entangled photons through the biexciton cascade decay in a single semiconductor quantum dot. To accomplish a high degree of entanglement, despite the exciton finestructure splitting, one must either energetically align the two exciton states by means of external fields or erase the which-path information by post-selecting photons within the correct frequency range. Here we show that in the latter scheme, as well as in related proposals, the photon state becomes deteriorated through dephasing processes in the solid, and the degree of entanglement remains low despite filtering.

\end{abstract}


% 73.21.La Quantum dots
% 03.67.Mn Entanglement measures, witnesses, and other characterizations
% 71.35.-y Excitons and related phenomena

\maketitle

 %%%%%%%%%%%%%%%%%%%%%%
 %%%  INTRODUCTION  %%%
 %%%%%%%%%%%%%%%%%%%%%%

\section{Introduction}

Conventional optics uses lenses and filters in order to manipulate light. The same elements are used for the manipulation of single or entangled photons.~\cite{bachor:98,leonhardt:03} Indeed, in most cases of interest the working principle of passive optical elements is identical at the classical and quantum level, but there are situations were things are more cumbersome. Spectral filtering at the single photon level, for instance, affects the photon wavepacket and therefore modifies the quantum state of the photon. This has important consequences for the creation of entangled photon pairs in single semiconductor quantum dots,~\cite{benson:00} which has recently been demonstrated~\cite{akopian:06,stevenson:06} and plays a key role in quantum information applications.~\cite{knill:01,gisin:02}

In semiconductor quantum dots electrons and holes can be trapped to form excitonic complexes, which consecutively decay through photon emission. Because of Coulomb correlation effects, the excitonic states usually have different energies and thus can be discriminated in optical experiments.~\cite{bayer:00a} Quantum-dot based photon devices generally use the last or last two photons emitted in such a cascade decay.~\cite{gerard:99,michler:00,benson:00} The last photon originates from the decay of the exciton, consisting of one electron and hole in the respective lowest unoccupied state, and the last but one photon from the decay of the biexciton state, consisting of two excitons with opposite spin orientations. The ambiguity of the biexciton decay, through either of the two ideally spin-degenerate exciton states, translates to a polarization entanglement of the emitted photons, that has been measured recently.~\cite{akopian:06,stevenson:06}

The degree of entanglement is largely diminished by the electron-hole exchange interaction,~\cite{bayer:02} which splits the energies of the intermediate exciton states by a small amount and attaches a which-path information to the photon frequencies. Several experimental implementations and proposals exist for overcoming this deficiency. First, the states can be brought back to degeneracy by means of external magnetic~\cite{stevenson:06,young:06,hudson:07} or electric~\cite{gerardot:07} fields, as well as through cavity modes with sufficiently strong coupling to both exciton and biexciton transitions.~\cite{stace:03} Alternatively, spectral filtering can be used to post-select only those photons whose energy contains no which-path information.~\cite{akopian:06} Finally, motivated by experimental observations of quantum dots with a vanishing biexciton binding energy, it has been proposed to entangle photons produced in different generations of the decay process.~\cite{reimer:07,avron:08} As we will show in this paper, only the first protocol provides a viable means for reaching a high degree of entanglement. The other two inherently suffer from dephasing processes in the solid, which are always significant, and the degree of entanglement will consequently remain rather low, thus limiting the performance of single-photon devices.~\cite{santori:02b,santori:04} Our results suggest that protocols for solid-state based quantum cryptography are more strict than previously thought.

% Outline
In this paper we provide analytical results for the degree of entanglement for the photons generated in the biexciton cascade decay in semiconductor quantum dots. We compare the different protocols based on spectral alignment of the intermediate exciton state, postselection through spectral filtering, and time reordering of photons produced in different generations. Our paper has been organized as follows. In Sec.~\ref{sec:theory} we introduce our theoretical model, and show how to compute the two-photon density matrix by means of the quantum regression theorem. Section~\ref{sec:results} presents the results obtained for the diffrent protocols and for realistic quantum dot parameters. Finally, in Sec.~\ref{sec:conclusions} we draw some conclusions.

 %%%%%%%%%%%%%%%%%%%%%%%%%%%
 %%%  THEORETICAL MODEL  %%%
 %%%%%%%%%%%%%%%%%%%%%%%%%%%

\section{Theory} \label{sec:theory}

 \begin{figure}
%BoundingBox 40 80 590 780
\includegraphics[width=0.7\columnwidth,angle=270]{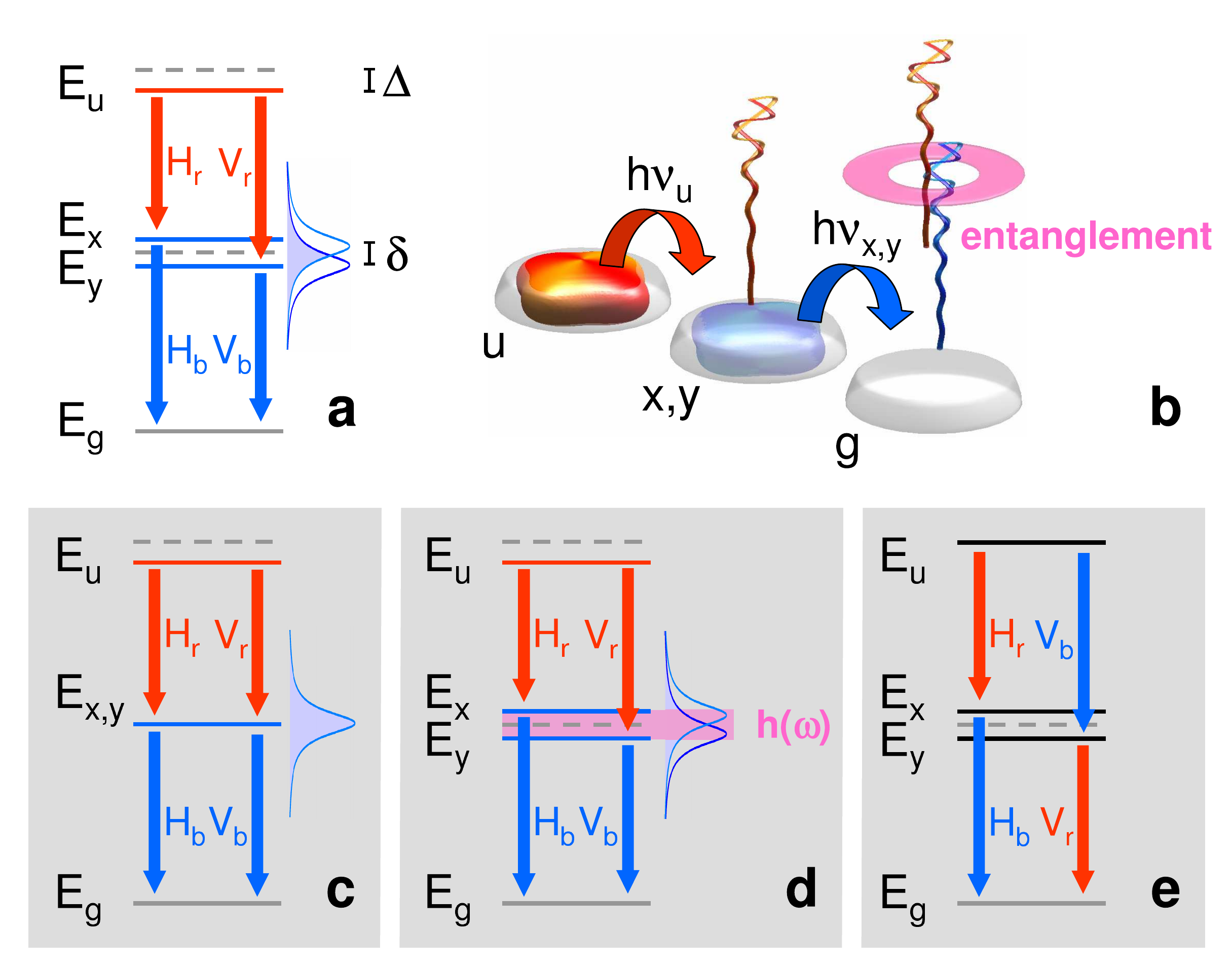}
\caption{(Color online) (a) Level scheme consisting of biexciton state $u$, exciton states polarized along $x$ and $y$, and groundstate $g$. The biexciton energy is reduced by the biexciton binding $\Delta$ and the exciton states are energetically separated  by the finestructure splitting $\delta$. $H$ and $V$ denote the polarizations of the emitted photons. (b) Schematic of creation of entangled photons. The biexciton first decays by emitting a photon with polarization $H$ or $V$, leaving behind a single exciton $x$ or $y$ in the dot, which decays in a second step. The ambiguity of the biexciton decay, through either of the two ideally spin-degenerate exciton states, translates to an entanglement of the emitted photons.~\cite{benson:00} Because of the finestructure splitting, the degree of entanglement is largely diminished. Several experimental implementations and proposals exist for overcoming this deficiency: (c) the states can be brought back to degeneracy by means of external fields,~\cite{stevenson:06} (d) spectral filtering can be used to post-select only those photons whose energy contains no which-path information,~\cite{akopian:06} with $h(\omega)$ being the filter function, or (e) photons of different generations of the cascade can be entangled by bringing the biexciton transition in resonance with the exciton one.~\cite{reimer:07,avron:08} \label{fig:schematics}
}\end{figure}

\subsection{Quantum state tomography} \label{sec:tomography}
 
In our theoretical approach, we consider the quantum dot level scheme depicted in Fig.~1a, consisting of the biexciton state $u$, the two exciton states with polarizations along $x$ and $y$, and the ground state $g$. By optical pumping~\cite{michler:00,akopian:06,stevenson:06} or electrical injection of carriers~\cite{benson:00,yuan:02} the system is initially prepared in the biexciton state $u$. Thereafter it decays radiatively in a cascade process by emitting two photons. Because of the biexciton binding $\Delta$, typically of the order of a few meV, the two photons have different energies and can be spectrally discriminated. We shall refer to the photons from the biexciton and exciton decay as red and blue, respectively. If the intermediate exciton states are degenerate, the two decay paths are indistinguishable and the photons become entangled in polarization.~\cite{benson:00} Such ideal performance is spoiled by the electron-hole exchange splitting,~\cite{bayer:02} as indicated in Fig.~\ref{fig:schematics}. To quantify the degree of entanglement, one usually performs a quantum state tomography~\cite{akopian:06,stevenson:06} by measuring the correlation between the two photons with a selected polarization combination, and averaging over the photon arrival times. This procedure provides us with the two-photon density matrix~\cite{mandel:95,troiani:06b}
\begin{equation}\label{eq:twophoton:I}
  \rho^{(2)}=\mbox{avg}\left[\, \left<:\!\hat I(t_r) \hat I(t_b)\!: \right> \right]
\end{equation}
which depends on the photon intensities $\hat I(t)$ of the first photon at time $t_r$ and the second photon at time $t_b$. The elements of $\rho^{(2)}$
\begin{equation} \label{eq:twophoton}
  \rho_{\lambda_r\lambda_b,\mu_r\mu_b}^{(2)} = \mbox{avg}\left[\, \left< 
  \hat E_{\lambda_r}^-(t_r)\hat E_{\lambda_b}^-(t_b)
  \hat E_{\mu_b}^+(t_b)\hat E_{\mu_r}^+(t_r)\right> \right]\,,
\end{equation}
then allow a complete reconstruction of the two-photon state. $\lambda$ and $\mu$ denote the horizontal and vertical photon polarizations $H$ and $V$, respectively. $\hat E_\lambda^\pm$ is the electric field operator for polarization $\lambda$ and for a positive or negative frequency component.~\cite{mandel:95}
A schematic picture of this procedure is depicted in Fig. \ref{fig7}(a) for the diagonal elements of the two-photon density matrix, and in Fig. \ref{fig7}(b) for the off-diagonal elements.

For the biexciton cascade decay shown in Fig.~1 and in absence of exciton spin scatterings,~\cite{akopian:06,stevenson:06} the only non-zero elements of the two-photon density matrix are $\rho_{H\!H,H\!H}^{(2)}$ and $\rho_{H\!H,V\!V}^{(2)}$, and those obtained by interchanging $H$ with $V$. The entanglement of formation or concurrence,~\cite{wooters:98} which provides a quantitative measure of the entanglement of the two emitted photons, is then given by the ratio of the off-diagonal and diagonal elements of $\rho^{(2)}$ as
\begin{equation}\label{eq:C}
C=\left|\frac{\rho^{(2)}_{H\!H,V\!V}} {\rho^{(2)}_{H\!H,H\!H}}\right| \,.
\end{equation}
The concurrence is one for perfect entanglement and zero in case of no entanglement.

\subsubsection{Unfiltered case}

To establish a relation between the electric field operators $\hat E^\pm$ and the quantum dot dynamics, we employ the Green function $G$ of classical Maxwell theory, which gives the field distribution for a localized current source. Let us consider the unfiltered case first. The expression~\cite{jackson:99} 
\begin{equation}\label{eq:G}
\hat E_\lambda^\pm(t+t_{\rm delay})=k^2G\,\hat d_\lambda^\pm(t)
\end{equation}
then connects the electric field with the dipole operator $\hat d_\lambda^\pm(t)$ of the excitonic transitions in the quantum dot. $k$ is the wavenumber in vacuum. We have assumed that the photons are emitted with a well-defined polarization $\lambda$ along the growth direction, as is the case for quantum dots placed inside a microcavity.~\cite{benson:00,michler:00} The main effect of the Green function is to introduce a time delay $t_{\rm delay}$ between the photon emission and its detection. We can now relate the photon-photon correlation function in Eq. \eqref{eq:twophoton} to a dipole-dipole correlation function
\begin{equation}\label{eq:twodipole}
  \mathcal{G}_{\lambda_r\lambda_b,\mu_r\mu_b}= \left< 
  \hat d_{\lambda_r}^-(t_r)\hat d_{\lambda_b}^-(t_b)
  \hat d_{\mu_b}^+(t_b)\hat d_{\mu_r}^+(t_r)\right> \,,
\end{equation}
where the dipole operators $\hat d_{\lambda_c}^\pm$ describe transitions between different exciton states, associated with creation or destruction of a photon with given polarization $\lambda$ and color $c$. For simplicity, we have not indicated the constant time delay $t_{\rm delay}$ between photon emission and detection. To arrive at the two-photon density matrix of Eq.~\eqref{eq:twophoton}, we have to multiply $\mathcal{G}$ with the Green functions of Maxwell's theory, which describe the propagation of the photon fields from the dot to the photon detector, and to integrate over the photon detection times $t_r$ and $t_b$. Note that the multiplication of $\mathcal{G}$ with the Green functions $G$ results in a mere constant, which drops out when calculating the two-photon density matrix or concurrence owing to the normalization condition ${\rm tr}[\rho^{(2)}]=1$.

\subsubsection{Filtered case}

\begin{figure}[t]
\includegraphics[width=0.8\columnwidth]{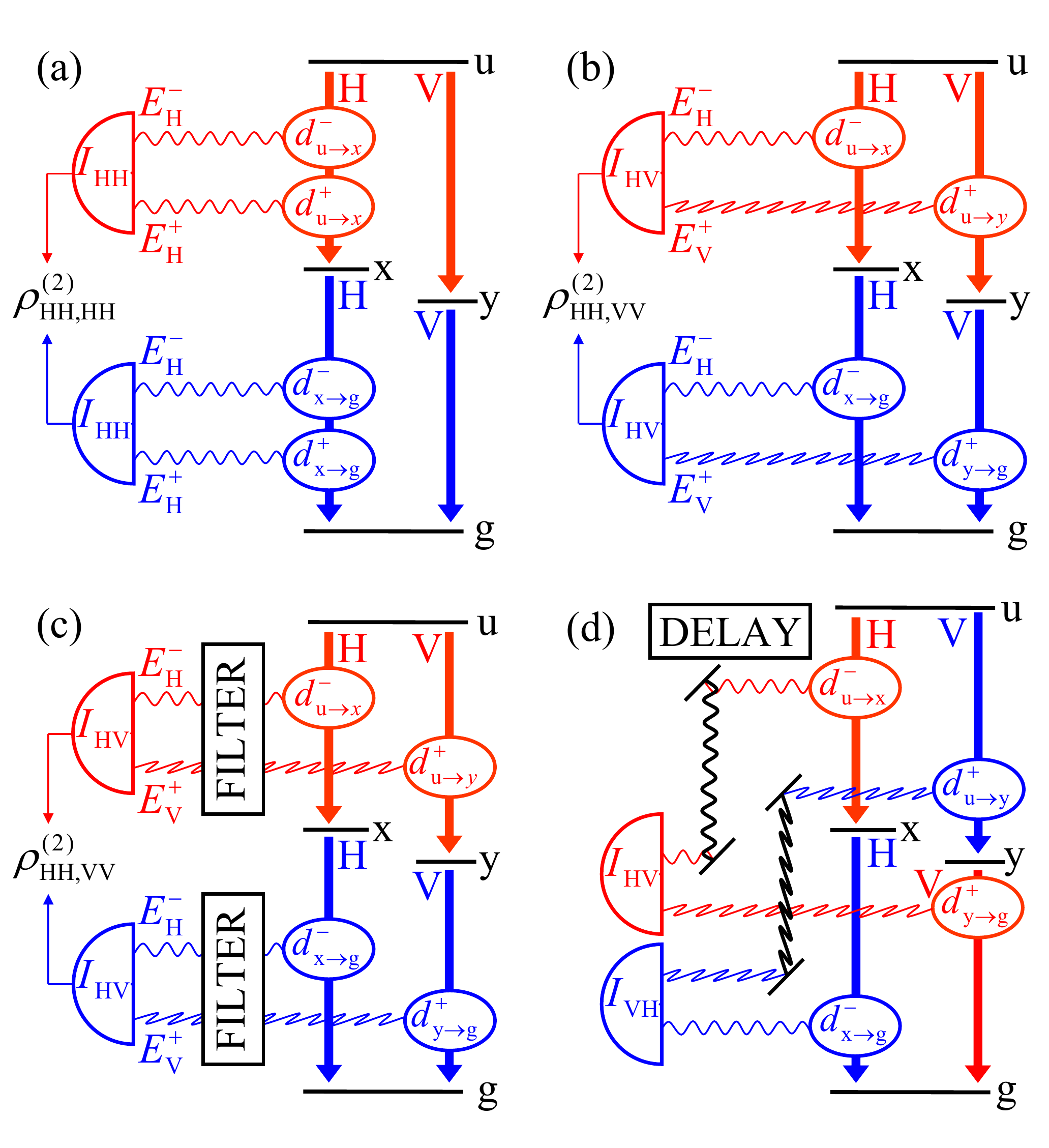}
\caption{(Color online) 
Schematic of relation between quantum dot cascade decay and two-photon density matrix.
(a) Diagonal element of the two-photon density matrix $\rho^{(2)}_{H\!H,H\!H}$ resulting from the decay of the biexciton $u$ into the $x$ polarized exciton and subsequently into the ground state $g$. The dipole operators $d$ mediate the electric field amplitudes $E$, as described by Eq. \eqref{eq:G}. We have indicated the states involved in the dipole transition rather than the polarization and color associated with the emitted photon. Wiggled lines indicate emitted photons, with polarization $H$ or $V$ depending on the orientation. Two electric field amplitudes build up a photon intensity $I_{H\!H}$ at the detector (half circles). The two-photon density matrix is the correlation between measurements at the two detectors, as described in Eq.~\eqref{eq:twophoton:I}. The diagonal element $\rho^{(2)}_{H\!H,H\!H}$ is the result of averaging over the arrival times of the intensity $I_{H\!H}$ of the first (red) and second (blue) photon. (b) Off-diagonal element of the two-photon density matrix $\rho^{(2)}_{H\!H,V\!V}$ resulting from the decay of the biexciton $u$ into a superposition of the $x$ and $y$ excitons. (c) $\rho^{(2)}_{H\!H,V\!V}$ with a filter modifying the electric field amplitudes arriving at the detectors, according to Eq.~\eqref{eq:filter}. (d) Time reordering by delaying the first generation of photons by a constant time, indicated by the thick black line between mirrors.
}
\label{fig7} \end{figure}

In the filtered case, the electric field is obtained from the expression
\begin{equation}\label{eq:filter}
  \hat E^+_\lambda(t+t_{\rm delay})=\int_0^t h(t-\tau)\,k^2G\,
  \hat d^+_\lambda(\tau)\,d\tau\,,
\end{equation}
with $h$ being the filter function. Upon Fourier transformation, the convolution becomes the familiar product $h(\omega)\hat d^+_\lambda(\omega)$ of filter function and signal source. The terms under the integral describe how the photon wavepacket, created through excitonic decay, propagates away from the dot and becomes modulated by the filter [see Fig. \ref{fig7}(c)]. Due to the filter process, where the frequency information is extracted from the temporal field evolution, $\rho^{(2)}$ is determined by a correlation function, similar to Eq.~\eqref{eq:twodipole}, but now with all four dipole operators acting at different times
\begin{widetext}
\begin{eqnarray}\label{eq:twodipolefilter}
  \mathcal{G}_{\lambda_r\lambda_b,\mu_r\mu_b}&=&
  \int_0^{t_r} h^*(t_r-\tau_r)\,d\tau_r
  \int_0^{t_b} h^*(t_b-\tau_b)\,d\tau_b
  \int_0^{t_b} h(t_b-\tau_b')\,d\tau_b'
  \int_0^{t_r} h(t_r-\tau_r')\,d\tau_r' \nonumber\\ &\times &
  \left< 
  \hat d_{\lambda_r}^-(\tau_r)\hat d_{\lambda_b}^-(\tau_b)
  \hat d_{\mu_b}^+(\tau_b')\hat d_{\mu_r}^+(\tau_r')\right>\,.
\end{eqnarray}
\end{widetext}
For a broadband filter, whose spectral width is large in comparison to the excitonic linewidth, the filter function becomes delta-like in time, and one recovers the dipole correlation function \eqref{eq:twodipole} of the unfiltered case. On the other hand, for a narrowband filter the integrals extend over a considerable time interval of the emission process.

\subsection{Quantum dot dynamics} \label{sec:dipole}

%\subsubsection{Quantum dot dynamics}

Next we show how to compute the dipole correlation function of Eqs.~\eqref{eq:twodipole} and \eqref{eq:twodipolefilter}. The quantum dot dynamics is described as an open quantum system interacting with the environment. The free propagation of the quantum dot states is governed by the Hamiltonian
\begin{equation}\label{eq:hamiltonian}
  \hat H=\sum_{i=x,y}E_i\,|i\rangle\langle i|+E_u\,|u\rangle\langle u|\,,
\end{equation}
\begin{table}[b]
\caption{ List of Lindblad operators used in our calculations. $\gamma_r$ and $\gamma_d$ denote the radiative and dephasing rates, respectively. We assume that the biexciton rates are two times larger than the exciton ones, which is a reasonable approximation for both radiative decay and pure dephasing.~\cite{hohenester.jpb:07} The operators for cross dephasing~\cite{hohenester.prl:07,hudson:07} and exciton spin scatterings are given in the last two lines for completeness, but are not considered in the calculations given in Sec.~\ref{sec:theory} [except for Eq.~\eqref{eq:rhocond}]. $i$ labels the two exciton states $x$ and $y$.}\label{table:lindblad}
\begin{ruledtabular} \begin{tabular}{llc} 
Description & Lindblad operator & Considered \\ \hline 
Radiative exciton decay & $\phantom{2}\sqrt{\gamma_r}\,|0\rangle\langle i|$ & yes \\
Radiative biexciton decay & $\sqrt{2\gamma_r}\,|i\rangle\langle u|$ & yes \\
Exciton dephasing & 
$\phantom 2\sqrt{\gamma_d}\,\left(|x\rangle\langle x|+|y\rangle\langle y|\right)$ & yes \\
Biexciton dephasing & $\sqrt{2\gamma_d}\,|u\rangle\langle u|$ & yes \smallskip \\
Cross dephasing & $\phantom 2\sqrt{\gamma_1}\,|i\rangle\langle i|$ & no \\
Exciton spin scattering & $\phantom 2\sqrt{\gamma_2}\,|y\rangle\langle x|$,
$\sqrt{\gamma_2}\,|x\rangle\langle y|$  & no \\
\end{tabular} \end{ruledtabular} \end{table}
with $i$ denoting the two different exciton states $x$ and $y$ with energy $E_x$ and $E_y$, respectively, and $u$ the biexciton state with energy $E_u$. Interaction with the environment results in transitions between the different quantum dot states, and dephasing. We consider radiative decay and pure dephasing, which are expected to be the main scattering channels for excitons in quantum dots. Pure dephasing is due to phonon couplings \cite{borri:01} and spectral diffusion due to charging centers in the vicinity of the dot. Other scattering channels, such as cross dephasing~\cite{hohenester.prl:07,hudson:07} or exciton spin scatterings, are usually of minor importance, and will be discussed at the end. We use the most simple description of radiative decay and pure dephasing in terms of scattering and dephasing rates $\gamma_r$ and $\gamma_d$, respectively. This will allow us to calculate the elements of $\rho^{(2)}$ analytically. Our rate equation approach is valid under the same conditions as the Born-Markov approximation,~\cite{breuer:02} namely that the memory kernel describing the buildup of scatterings is sufficiently short in comparison to the scattering times. The memory of phonon dephasing, which is expected to be the slowest process, is of the order of a few picoseconds~\cite{krummheuer:02,hohenester.jpb:07} and thus much shorter than the dephasing time $\gamma_d^{-1}$ of several tens to hundreds of picoseconds. Quite generally, it is known that phonon dephasing cannot be fully described by a simple dephasing rate, and the lineshape of excitonic transitions consists of a non-Lorentzian phonon background and a zero-phonon line.~\cite{borri:01,krummheuer:02} The dephasing rate $\gamma_d$ of our work mimics an averaged decay characteristics. We employ a master equation approach of Lindblad form~\cite{walls:95}
\begin{equation}\label{eq:master}
  i\dot{\hat\rho}=[\hat H,\hat\rho]-\frac i2\sum_\mu
  \left(\hat L_\mu^\dagger\hat L_\mu^{\phantom\dagger}\,\hat\rho+
  \hat\rho\,\hat L_\mu^\dagger\hat L_\mu^{\phantom\dagger}-2
  \hat L_\mu^{\phantom\dagger}\hat\rho\hat L_\mu^\dagger
  \right)\,,
\end{equation}
where $\hat\rho$ is the density operator of the quantum dot states, and the Lindblad operators $\hat L_\mu$ describe the various scattering channels listed in Table~\ref{table:lindblad}. For the Lindblad operator associated with radiative decay, the last term in the parentheses of Eq.~\eqref{eq:master} corresponds to the excitonic decay process, within which a photon is emitted from the dot.

%\subsubsection{Quantum regression theorem}

Introducing the shorthand notation $\mathcal{L}\hat\rho$ for the right-hand side of Eq.~\eqref{eq:master}, where $\mathcal{L}$ is the Liouville superoperator,~\cite{breuer:02} we can write the formal solution of the master equation in the form
\begin{equation}\label{eq:mastersolution}
  \hat\rho(t)=e^{-i\mathcal{L}t}\hat\rho_0=
  \left\{\left(e^{-i\mathcal{L}t}\right)_{\rm no\,\,em}+
  \left(e^{-i\mathcal{L}t}\right)_{\rm em}\right\}\hat\rho_0\,,
\end{equation}
with $\hat\rho_0$ being the initial density operator. On the right-hand side of Eq.~\eqref{eq:mastersolution} we have split the time evolution operator into two contributions, the first one associated with no photon emission, which does not change the number of excitons, and the second one associated with photon emission, which changes the number of excitons. It will turn out that only the first part is needed in the calculation of the dipole correlation function \eqref{eq:twodipole}, which will allow us to considerably simplify our analysis.

We next briefly discuss how to compute multi-time correlation functions of the form $\langle\hat A_1(t_1)\hat A_2(t_2)\dots\rangle$, such as the dipole-dipole correlation function, where $\hat A$ is an arbitrary operator. This is conveniently accomplished by the {\em quantum regression theorem},~\cite{gardiner:04,mandel:95} which allows to express multi-time correlation functions in terms of single-time expectations. It implies that the fluctuations regress in time like the macroscopic averages, and holds exactly for systems initially decoupled from the environment and subject to a Markovian dynamics. Both approximations are valid for the description of the photon cascade decay in a quantum dot. For the initial decoupling it suffices to assume that the system will pass, at some point of the cascade, through the biexciton state, which consecutively decays through photon emissions. The applicability of the Markov approximation has been justified before in Sec.~\ref{sec:dipole}. According to the quantum regression theorem, a multi-time correlation function can be unraveled to a sequence of time evolutions, governed by $e^{-i\mathcal{L}t}$, which are interrupted by abrupt transitions at times $t_1<t_2<\dots$ according to~\cite{gardiner:04}
\begin{equation}\label{eq:regression}
  \langle\hat A_1(t_1)\hat A_2(t_2)\dots\rangle = \mbox{tr}\left[\,
  \dots\, \mathcal{A}_2 e^{-i\mathcal{L}(t_2-t_1)}\,
          \mathcal{A}_1 e^{-i\mathcal{L}t_1}\hat\rho_0\right]\,.
\end{equation}
Here $\mathcal{A}\hat\rho$ is a short-hand notation for $\hat A\hat\rho$, if $\hat A$ is an operator evolving with positive frequency such as $\hat d^+$, and for $\hat\rho\hat A$, if $\hat A$ is an operator evolving with negative frequency such as $\hat d^-$.

There is a subtle point regarding the use of the quantum regression theorem for the calculation of the dipole correlation function of Eqs.~\eqref{eq:twodipole} and \eqref{eq:twodipolefilter} in case of a biexciton cascade, where the system is initially in the biexciton state $\hat\rho_0=|u\rangle\langle u|$. According to Eq.~\eqref{eq:regression}, the dipole operators $\hat d^\pm$ act from either the left or right hand side on the time-propagated $\hat\rho_0$, each of them creating a photon field. As we can create at most two photon fields from either side of $\hat\rho_0$, it immediately follows that we have to consider in the time evolutions in Eq.~\eqref{eq:regression} only that part of $e^{-i\mathcal{L}t}$ associated with no photon emission. The remainder corresponds to terms where more than four photon fields are involved, which are automatically zero in case of a pure cascade decay.

All Lindblad operators considered in our calculations (see Table~\ref{table:lindblad}) either correspond to pure dephasing, which does not mix the different excitonic states, or to radiative decay. Thus, if the system is initially in the state $|p\rangle\langle q|$, with $p$ and $q$ labeling the quantum dot states $g$, $x$, $y$, and $u$, the conditional time evolution $\left(e^{-i\mathcal{L}t}\right)_{\rm no\,\,em}$ does not couple to any other state. From the master equation \eqref{eq:master} we then obtain for the conditional time evolution within the usual rotating wave approximation~\cite{walls:95}
\begin{widetext}%\scriptsize
\begin{eqnarray}\label{eq:rhocond}
  &&\rho_{pq}=\left(e^{-i\mathcal{L}t}\right)_{\rm no\,\,em}|p\rangle\langle q|=\nonumber\\
  &&\quad{\scriptsize\left(\begin{array}{llll}
    1 & 
    \exp\!\!\left\lgroup-i\frac\delta 2 t-\frac{\gamma_{\rm tot}}2 t\right\rgroup &
    \exp\!\!\left\lgroup-i\frac\delta 2 t-\frac{\gamma_{\rm tot}}2 t\right\rgroup &
    \exp\!\!\left\lgroup\phantom-i\Delta t-\gamma_{\rm tot} t\right\rgroup \\
    \exp\!\!\left\lgroup\phantom-i\frac\delta 2 t-\frac{\gamma_{\rm tot}}2 t\right\rgroup &
    \exp\!\!\left\lgroup-\gamma_r t\right\rgroup &
    \exp\!\!\left\lgroup-i\delta t-(\gamma_r+\gamma_1) t\right\rgroup &
    \exp\!\!\left\lgroup\phantom-i(\Delta-\frac\delta 2) t-\frac{3\gamma_{\rm tot}}2 t\right\rgroup \\
    \exp\!\!\left\lgroup\phantom-i\frac\delta 2 t-\frac{\gamma_{\rm tot}}2 t\right\rgroup &
    \exp\!\!\left\lgroup\phantom-i\delta t-(\gamma_r+\gamma_1) t\right\rgroup &
    \exp\!\!\left\lgroup-\gamma_r t\right\rgroup &
    \exp\!\!\left\lgroup\phantom-i(\Delta+\frac\delta 2) t-\frac{3\gamma_{\rm tot}}2 t\right\rgroup \\
    \exp\!\!\left\lgroup-i\Delta t-\gamma_{\rm tot} t\right\rgroup &
    \exp\!\!\left\lgroup-i(\Delta-\frac\delta 2) t-\frac{3\gamma_{\rm tot}}2 t\right\rgroup &
    \exp\!\!\left\lgroup-i(\Delta+\frac\delta 2) t-\frac{3\gamma_{\rm tot}}2 t\right\rgroup &
    \exp\!\!\left\lgroup-2\gamma_r t\right\rgroup \\
  \end{array}\right)_{pq}}\,.\nonumber\\
\end{eqnarray}
\end{widetext}
where $\gamma_{\rm tot}=\gamma_r+\gamma_d$ is the sum of radiative and dephasing rates, $\delta$ is the finestructure splitting, and $\Delta$ is the biexciton binding energy. The components of the conditional density matrix have one part associated with the free time evolution due to $\hat H$, and another part associated with scatterings. The imaginary parts of the arguments of the exponentials are due to the different energies of the excitonic states, and the real parts due to scattering processes. Pure dephasing only affects those elements $\rho_{pq}$ which have an unequal number of excitons in $p$ and $q$. We have included cross dephasing~\cite{hohenester.prl:07,hudson:07} $\gamma_1$ in Eq.~\eqref{eq:rhocond}, which affects the elements $\rho_{xy}$ and $\rho_{yx}$. This dephasing mechanism will not be explicitly considered in our following calculations, but will be addressed in our final discussion.

\section{Results}\label{sec:results}

\subsection{Unfiltered case}

We now have all ingredients at hand to compute the elements of the two-photon density matrix. Let us consider the unfiltered case first. Following the prescription of the quantum regression theorem, Eq.~\eqref{eq:regression}, the correlation function is unraveled to a sequence of conditional density matrices $\rho_{uu}\to\rho_{\lambda\mu}\to\rho_{gg}$, which evolve under the condition that the system undergoes abrupt transitions, determined by the sequence of dipole operators, at times $t_r$ and $t_b$. From the elements of the conditional density matrix of Eq.~\eqref{eq:rhocond}, we get for the diagonal and off-diagonal elements of the dipole-dipole correlation function
\begin{eqnarray}\label{eq:twodipoletime}
  \mathcal{G}_{H\!H,H\!H}&=&\mathcal{N}\,e^{-2\gamma_r t_r}e^{-\gamma_r(t_b-t_r)}
  \nonumber\\
  \mathcal{G}_{H\!H,V\!V}&=&\mathcal{N}\,e^{-2\gamma_r t_r}
  e^{-\gamma_r(t_b-t_r)-i\delta(t_b-t_r)}\,,
  \nonumber\\
\end{eqnarray}
where $\mathcal{N}$ is a normalization constant that includes dipole moments of different excitonic transitions. The entanglement of formation or concurrence [Eq. \eqref{eq:C}] is obtained by averaging in Eq.~\eqref{eq:twodipoletime} over the photon arrival times $t_r$ and $t_b$,
\begin{equation}\label{eq:concurrence}
  C=\left|\frac{\gamma_r}{\gamma_r+i\delta}\right|\,.
\end{equation}
It has its maximum for zero finestructure splitting, and decreases with increasing $\delta$, as shown in Fig.~3. This is because of the different phases acquired in the evolution of the intermediate exciton state, and the time average over the resulting phase factors.

\begin{figure}[b]
\includegraphics[width=0.82\columnwidth]{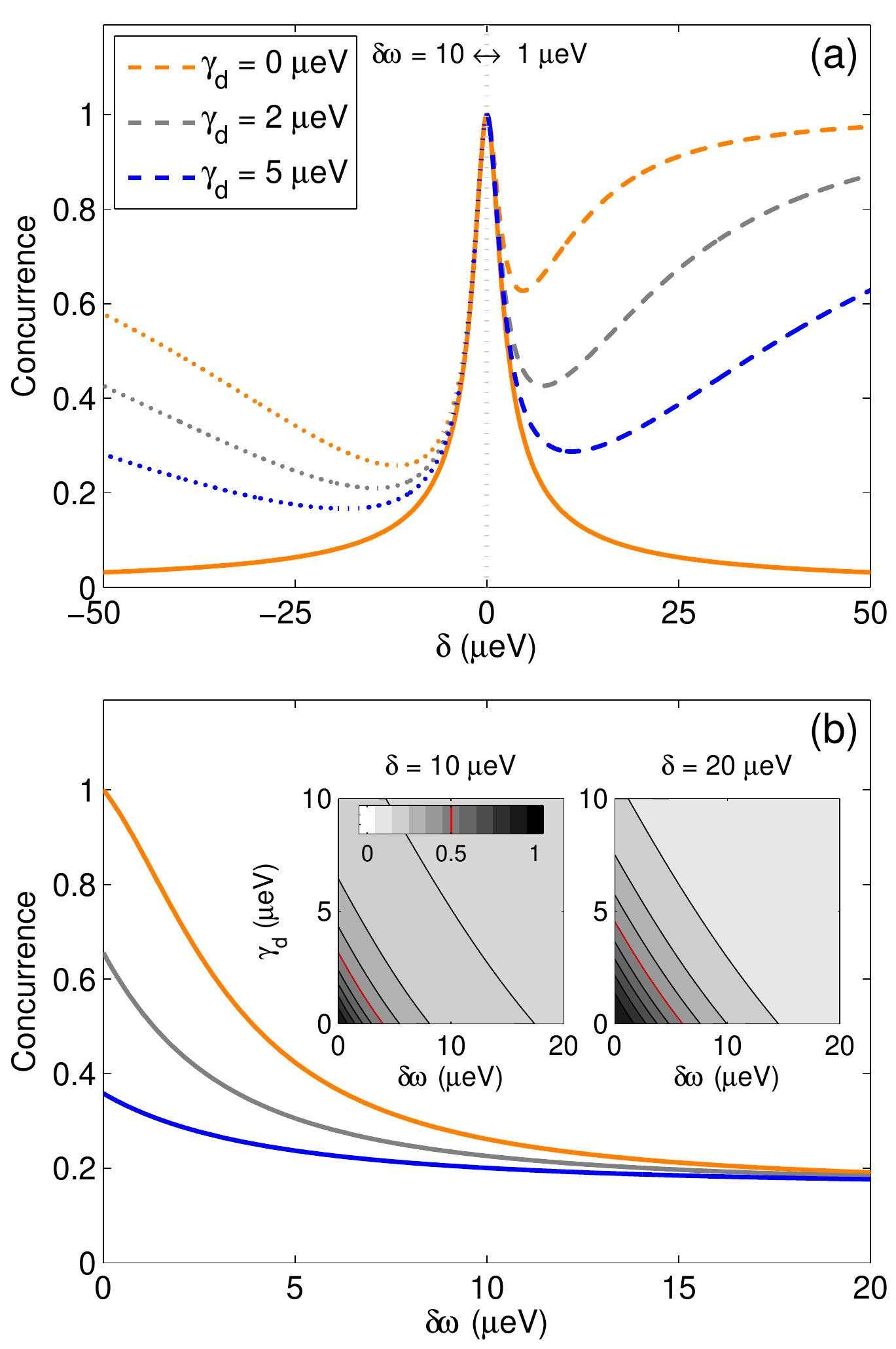}
\caption{(Color online) (a) Concurrence as a function of finestructure splitting and for different dephasing rates, as computed from the expressions given in table~\ref{table:concurrence}. The solid line corresponds to the unfiltered case. The dashed (dotted) lines report results for $\delta\omega=1\,\mu$eV ($10\,\mu$eV) which are plotted for positive (negative) finestructure splittings only. We use a radiative decay rate of $\gamma_r=1.6\,\mu$eV which corresponds to a lifetime of about one nanosecond. (b) Concurrence as a function of filter width $\delta\omega$ and for $\delta=10\,\mu$eV. In presence of dephasing the concurrence remains low despite filtering. The insets report density maps of the concurrence as a function of filter width and dephasing rate. 
}
\end{figure}

%%%%%%%%%%%%%%%%%%%%%%%%%%%%%%%%%%%%%%%%%%%%%%%%%%%%%%%%%%%%%%%%%%%%%%%%%%%%%%%%%%%%%%
\begin{table*}
\caption{ Elements of the two-photon density matrix and details of their calculation by means of the quantum regression theorem.~\cite{mandel:95,gardiner:04} In absence of exciton spin scatterings, the only nonzero elements are $\rho^{(2)}_{H\!H,H\!H}$ and $\rho^{(2)}_{H\!H,V\!V}$, and those obtained by interchanging $H$ and $V$. The concurrence~\cite{wooters:98} is given by the modulus of the ratio between the off-diagonal and diagonal element.~\cite{akopian:06,stevenson:06} The correlation function is unraveled to a sequence of conditional density matrices, which evolve under the condition that the system undergoes abrupt transitions. Note that $\rho^{(2)}$ is not properly normalized in case of filtering due to the loss of photons. In the entanglement of photons in different generations,~\cite{reimer:07,avron:08} we assume that the photons originating from the biexciton decay are delayed by some time $t_0$. We use $\gamma_{\rm tot}=\gamma_r+\gamma_d$ and $\mbox{c.c.}$ for the complex conjugate of the preceding term.
}\label{table:concurrence}

\begin{ruledtabular}
\begin{tabular}{llllr} 
Figure & Basis $\lambda\mu$ & Time order & Decay path & $\rho^{(2)}_{\lambda\lambda,\mu\mu}$ \\
\hline
Fig.~1a &
$H\!H$ &
$0\le t_r\le t_b$ &
$\rho_{uu}(t_r)\to\rho_{xx}(t_b-t_r)\to\rho_{gg}$ &
$\displaystyle\frac 12$ \\
&
$HV$ &
$0\le t_r\le t_b$ &
$\rho_{uu}(t_r)\to\rho_{xy}(t_b-t_r)\to\rho_{gg}$ &
$\displaystyle{\frac 12\frac{\gamma_r}{\gamma_r+i\delta}}$ \medskip \\ 
Fig.~1d &
$H\!H$\footnote{Other decay paths are obtained through $\tau\leftrightarrow\tau'$ and $\rho_{pq}\to\rho_{qp}$. For the off-diagonal elements one has to additionally replace $x\leftrightarrow y$.} &
$0\le\tau_r\le\tau_r'\le\tau_b\le\tau_b'$ &
$\rho_{uu}(\tau_r)\to\rho_{xu}(\tau_r'-\tau_r)\to\rho_{xx}(\tau_b-\tau_r')
                                  \to\rho_{gx}(\tau_b'-\tau_b)\to\rho_{gg}$
\\
& & $0\le\tau_r'\le\tau_r\le\tau_b\le\tau_b'$ &
$\rho_{uu}(\tau_r')\to\rho_{ux}(\tau_r-\tau_r')\to\rho_{xx}(\tau_b-\tau_r)
                                  \to\rho_{gx}(\tau_b'-\tau_b)\to\rho_{gg}$
\\
& & $0\le\tau_r\le\tau_b\le\tau_r'\le\tau_b'$ &
$\rho_{uu}(\tau_r)\to\rho_{xu}(\tau_b-\tau_r)\to\rho_{gu}(\tau_r'-\tau_b)
                                  \to\rho_{gx}(\tau_b'-\tau_r')\to\rho_{gg}$
\\
& &
\multicolumn{3}{r}{$\displaystyle
  \frac{(\delta\omega)^2}{8}
  \frac 1{\delta\omega+i\delta+3\gamma_{\rm tot}}
  \left(
  \frac 1{\delta\omega-i\delta+\gamma_{\rm tot}}+
  \frac 1{\delta\omega+i\delta+\gamma_{\rm tot}}+
  \frac{\gamma_r}{(\delta\omega+\gamma_{\rm tot})
                  (\delta\omega-i\delta+\gamma_{\rm tot})}\right)+\mbox{c.c.}$} \medskip \\
&
$HV$\footnotemark[1] &
$0\le\tau_r\le\tau_r'\le\tau_b\le\tau_b'$ &

$\rho_{uu}(\tau_r)\to\rho_{xu}(\tau_r'-\tau_r)\to\rho_{xy}(\tau_b-\tau_r')
                                  \to\rho_{gy}(\tau_b'-\tau_b)\to\rho_{gg}$
\\
& & $0\le\tau_r'\le\tau_r\le\tau_b\le\tau_b'$ &
$\rho_{uu}(\tau_r')\to\rho_{uy}(\tau_r-\tau_r')\to\rho_{xy}(\tau_b-\tau_r)
                                  \to\rho_{gy}(\tau_b'-\tau_b)\to\rho_{gg}$
\\
& & $0\le\tau_r\le\tau_b\le\tau_r'\le\tau_b'$ &
$\rho_{uu}(\tau_r)\to\rho_{xu}(\tau_b-\tau_r)\to\rho_{gu}(\tau_r'-\tau_b)
                                  \to\rho_{gy}(\tau_b'-\tau_r')\to\rho_{gg}$
\\ \medskip
& & &
\multicolumn{2}{r}{$\displaystyle
  \frac{(\delta\omega)^2}{4}
  \frac{\gamma_r}{(\delta\omega+i\delta+3\gamma_{\rm tot})
                  (\delta\omega+i\delta+\gamma_{\rm tot})}
  \left(\frac 2{\gamma_r+i\delta}+\frac 1{\delta\omega+\gamma_{\rm tot}}
  \right)$}\\
Fig.~1e &
$H\!H$ &
$0\le t_r-t_0\le t_b$ &
$\rho_{uu}(t_r-t_0)\to\rho_{xx}(t_b-t_r+t_0)\to\rho_{gg}$ &
$\displaystyle\frac 12$ \\
&
$HV$ &
$t_0\le t_r\le t_b$ &
$\rho_{uu}(t_r-t_0)\to\rho_{xu}(t_b-t_r)\to\rho_{xy}(t_r-t_b+t_0)\to
                                      \rho_{gy}(t_b-t_r)\to\rho_{gg}$ \\
& &
$t_0\le t_b\le t_r$ &
$\rho_{uu}(t_b-t_0)\to\rho_{uy}(t_r-t_b)\to\rho_{xy}(t_b-t_r+t_0)\to
                                      \rho_{xg}(t_r-t_b)\to\rho_{gg}$ \\
& & &
\multicolumn{2}{r}{$\displaystyle
  \frac{\gamma_r}{\gamma_r+2\gamma_d}\left(
  e^{-\gamma_rt_0}-
  e^{-2\gamma_{\rm tot}t_0}\right)$}\\
\end{tabular}
\end{ruledtabular}
\end{table*}
%%%%%%%%%%%%%%%%%%%%%%%%%%%%%%%%%%%%%%%%%%%%%%%%%%%%%%%%%%%%%%%%%%%%%%%%%%%%%%%%%%%%%%

\subsection{Filtered case}

The calculation of the two-photon density matrix in case of filtering is very similar but more cumbersome. We again employ the quantum regression theorem for its calculation, but we now obtain conditional density matrix elements $\rho_{pq}$ with unequal number of excitons, whose time evolutions are affected by pure dephasing. To simplify our analysis, we assume that the filter function $h(\omega)$ has a Lorentzian shape \cite{Knoell}, with a central frequency positioned in the middle of the two finestructure-split emission peaks, and width $\delta\omega$. Within this approach we can perform all integrals analytically. In the evaluation of the dipole correlation function \eqref{eq:twodipolefilter} we have to consider all possible time orderings for $\tau$, subject to the conditions $\tau_r\le\tau_b$ and $\tau_r'\le\tau_b'$ which are a direct consequence of the cascade nature of the biexciton decay. Table~\ref{table:concurrence} provides details about the resulting six time orderings, and lists the resulting diagonal and off-diagonal element of the two-photon density matrix (see also Fig.~2).

Two limiting cases can be readily obtained from the elements of the filtered two-photon correlation function. First, for $\delta\omega\to\infty$ the concurrence reduces to that of the unfiltered case. Second, for zero finestructure splitting the concurrence becomes one. This perfect entanglement is due to the absence of any which-path information in the photon frequencies. In the general case, the concurrence depends on the quantum dot parameters $\delta$, $\gamma_r$, and $\gamma_d$, as well as on the filter width $\delta\omega$. Results for typical quantum dot and filter parameters are shown in Fig.~3. Panel (a) reports that in the filtered case the degree of entanglement first drops with increasing $\delta$, and then increases again for larger values of $\delta$. This is due to the competition between the opposite trends of increase of which-path information with increasing $\delta$, and the masking of which-path information through the filter. The latter process is of importance when the two lines become energetically separated, but it comes at the price of a strongly reduced efficiency since in this regime practically all photons are absorbed by the filter. From the figure we observe that the concurrence dramatically drops in presence of dephasing. This is also evident from panel (b) which reports the influence of the filter width on the concurrence. For finite values of $\gamma_d$ the degree of entanglement remains low despite filtering. This is due to the transition of the pure entangled photon state to a mixed, classically correlated photon state, which takes place in the process of filtering. 

The precise value of the dephasing rate depends on the detailed quantum dot parameters and on the coupling of the excitons to the solid state environment. The line width of approximately 50 $\mu$eV measured in experiment~\cite{akopian:06,stevenson:06} provides a hypothetical upper limit. This value is certainly too high, as it includes spectral diffusion which occurs on longer timescales due to fluctuating environment fields. These fluctuations affect the energies of the excitonic states but are not expected to have a drastic impact on the entanglement properties. For a lower bound of $\gamma_d$ we estimate a dephasing time of 340 ps, corresponding to a rate of approximately 5 $\mu$eV, which was obtained in single-photon interferometer measurements.~\cite{santori:02b,santori:04,note.akopian} It is apparent from Fig.~3 that for such values of $\gamma_d$ the concurrence is already restricted to small values.

\subsection{Time reordering}

Dephasing also plays an important role for entanglement of photons created in different generations of the biexciton cascade (see Fig.~1e), which has recently been proposed for quantum dots with a vanishing biexciton binding energy.~\cite{reimer:07,avron:08} Using the notation of Fig.~1(e), within this approach one entangles the red photon fields $H_r$ and $V_r$ with the blue photon fields $H_b$ and $V_b$. This is accomplished by delaying the fields $H_r$ and $V_b$ originating from the biexciton decay by a constant time $t_0$ [see Fig. \ref{fig7}(d)]. The elements of the dipole-dipole correlation function then read
\begin{eqnarray}\label{eq:twodipoledelay}
  \mathcal{G}_{H\!H,H\!H}&=& \left< 
  \hat d_{H_r}^-(t_r-t_0)\hat d_{H_b}^-(t_b)
  \hat d_{H_b}^+(t_b)\hat d_{H_r}^+(t_r-t_0)\right> \nonumber\\
  \mathcal{G}_{H\!H,V\!V}&=& \left< 
  \hat d_{H_r}^-(t_r-t_0)\hat d_{H_b}^-(t_b)
  \hat d_{V_b}^+(t_b-t_0)\hat d_{V_r}^+(t_r)\right>\,.
\end{eqnarray}
Table~\ref{table:concurrence} provides the results for the elements of the two-photon density matrix, which are obtained by evaluating Eq.~\eqref{eq:twodipoledelay} by means of the quantum regression theorem and averaging over the photon arrival times. Note that in this scheme the finestructure splitting $\delta$ plays a different role than in the previous schemes, as it is used to distinguish the red from the blue photons. Employing again the rotating wave approximation, we consider this by setting $\delta$ equal to zero in the conditional density matrices \eqref{eq:rhocond}. In experiments, $\delta$ should be sufficiently large in order to allow for an efficient filtering of the different photon colors. However, even in the case where filtering of the red and blue photon has no effect on the photon wavepacket, as assumed in the results given in the table, the time jitter of the photon emissions and dephasing significantly reduce the overall concurrence. Remarkably, even in absence of any dephasing the two photon wavepackets emitted in the cascade decay do not overlap completely, for any time delay $t_0$, and the concurrence is bound to values below one half.

 %%%%%%%%%%%%%%%%%%%%
 %%%  DISCUSSION  %%%
 %%%%%%%%%%%%%%%%%%%%
 
\section{Conclusions}\label{sec:conclusions}

The radiative biexciton cascade decay in single semiconductor quantum dots has the potential of being a source of triggered polarization-entangled photon pairs. Whether devices based on this mechanism will prove successful for quantum information applications depends on a number of issues. A high collection efficiency of photons through the microcavity is crucial. In single-photon devices radiative scattering enhancements by a factor of ten or more have been demonstrated,~\cite{santori:02b,santori:04} but things are more difficult in entangled-photon sources due to the different photon colors. From a more fundamental perspective, the issue of exciton finestructure splitting, which is a consequence of the general quantum principle of level repulsion, has to be addressed properly. As we have shown here, through filtering~\cite{akopian:06} or entanglement of photons produced in different generations~\cite{reimer:07,avron:08} the degree of entanglement becomes affected through dephasing losses. Such losses limit the device performance already at low temperatures, but make entanglement generation probably impossible at elevated temperatures. In contrast, in the unfiltered case dephasing is a much less critical issue, with cross dephasing constituting the major cause of entanglement loss,~\cite{hohenester.prl:07,hudson:07} and the devices could operate at higher temperatures.~\cite{hafenbrak:07} We thus believe that the {\em tour de force}\/ approach of reducing the finestructure splitting, by means of growth optimization or external fields, is the most promising one for reaching a high degree of photon entanglement.

%\bibliography{../longbib,info}

\begin{thebibliography}{34}
\expandafter\ifx\csname natexlab\endcsname\relax\def\natexlab#1{#1}\fi
\expandafter\ifx\csname bibnamefont\endcsname\relax
  \def\bibnamefont#1{#1}\fi
\expandafter\ifx\csname bibfnamefont\endcsname\relax
  \def\bibfnamefont#1{#1}\fi
\expandafter\ifx\csname citenamefont\endcsname\relax
  \def\citenamefont#1{#1}\fi
\expandafter\ifx\csname url\endcsname\relax
  \def\url#1{\texttt{#1}}\fi
\expandafter\ifx\csname urlprefix\endcsname\relax\def\urlprefix{URL }\fi
\providecommand{\bibinfo}[2]{#2}
\providecommand{\eprint}[2][]{\url{#2}}

\bibitem[{\citenamefont{Bachor}(1998)}]{bachor:98}
\bibinfo{author}{\bibfnamefont{H.-A.} \bibnamefont{Bachor}},
  \emph{\bibinfo{title}{A Guide to Experiments in Quantum Optics}}
  (\bibinfo{publisher}{Wiley}, \bibinfo{address}{Wienheim},
  \bibinfo{year}{1998}).

\bibitem[{\citenamefont{Leonhardt}(2003)}]{leonhardt:03}
\bibinfo{author}{\bibfnamefont{U.}~\bibnamefont{Leonhardt}},
  \bibinfo{journal}{Rep. Prog. Phys.} \textbf{\bibinfo{volume}{66}},
  \bibinfo{pages}{1207} (\bibinfo{year}{2003}).

\bibitem[{\citenamefont{Benson et~al.}(2000)\citenamefont{Benson, Santori,
  Pelton, and Yamamoto}}]{benson:00}
\bibinfo{author}{\bibfnamefont{O.}~\bibnamefont{Benson}},
  \bibinfo{author}{\bibfnamefont{C.}~\bibnamefont{Santori}},
  \bibinfo{author}{\bibfnamefont{M.}~\bibnamefont{Pelton}}, \bibnamefont{and}
  \bibinfo{author}{\bibfnamefont{Y.}~\bibnamefont{Yamamoto}},
  \bibinfo{journal}{Phys. Rev. Lett.} \textbf{\bibinfo{volume}{84}},
  \bibinfo{pages}{2513} (\bibinfo{year}{2000}).

\bibitem[{\citenamefont{Akopian et~al.}(2006)\citenamefont{Akopian, Lindner,
  Poem, Berlatzky, Avron, Gershoni, Gerardot, and Petroff}}]{akopian:06}
\bibinfo{author}{\bibfnamefont{N.}~\bibnamefont{Akopian}},
  \bibinfo{author}{\bibfnamefont{N.~H.} \bibnamefont{Lindner}},
  \bibinfo{author}{\bibfnamefont{E.}~\bibnamefont{Poem}},
  \bibinfo{author}{\bibfnamefont{Y.}~\bibnamefont{Berlatzky}},
  \bibinfo{author}{\bibfnamefont{J.}~\bibnamefont{Avron}},
  \bibinfo{author}{\bibfnamefont{D.}~\bibnamefont{Gershoni}},
  \bibinfo{author}{\bibfnamefont{B.~D.} \bibnamefont{Gerardot}},
  \bibnamefont{and} \bibinfo{author}{\bibfnamefont{P.~M.}
  \bibnamefont{Petroff}}, \bibinfo{journal}{Phys. Rev. Lett.}
  \textbf{\bibinfo{volume}{96}}, \bibinfo{eid}{130501} (\bibinfo{year}{2006}).

\bibitem[{\citenamefont{Stevenson et~al.}(2006)\citenamefont{Stevenson, Young,
  Atkinson, Cooper, Ritchie, and Shields}}]{stevenson:06}
\bibinfo{author}{\bibfnamefont{R.~M.} \bibnamefont{Stevenson}},
  \bibinfo{author}{\bibfnamefont{R.~J.} \bibnamefont{Young}},
  \bibinfo{author}{\bibfnamefont{P.}~\bibnamefont{Atkinson}},
  \bibinfo{author}{\bibfnamefont{K.}~\bibnamefont{Cooper}},
  \bibinfo{author}{\bibfnamefont{D.~A.} \bibnamefont{Ritchie}},
  \bibnamefont{and} \bibinfo{author}{\bibfnamefont{A.~J.}
  \bibnamefont{Shields}}, \bibinfo{journal}{Nature (London)}
  \textbf{\bibinfo{volume}{439}}, \bibinfo{pages}{179} (\bibinfo{year}{2006}).

\bibitem[{\citenamefont{Knill et~al.}(2001)\citenamefont{Knill, Laflamme, and
  Milburn}}]{knill:01}
\bibinfo{author}{\bibfnamefont{E.}~\bibnamefont{Knill}},
  \bibinfo{author}{\bibfnamefont{R.}~\bibnamefont{Laflamme}}, \bibnamefont{and}
  \bibinfo{author}{\bibfnamefont{G.~J.} \bibnamefont{Milburn}},
  \bibinfo{journal}{Nature} \textbf{\bibinfo{volume}{409}}, \bibinfo{pages}{46}
  (\bibinfo{year}{2001}).

\bibitem[{\citenamefont{Gisin et~al.}(2002)\citenamefont{Gisin, Ribordy,
  Tittel, and Zbinden}}]{gisin:02}
\bibinfo{author}{\bibfnamefont{N.}~\bibnamefont{Gisin}},
  \bibinfo{author}{\bibfnamefont{G.}~\bibnamefont{Ribordy}},
  \bibinfo{author}{\bibfnamefont{W.}~\bibnamefont{Tittel}}, \bibnamefont{and}
  \bibinfo{author}{\bibfnamefont{H.}~\bibnamefont{Zbinden}},
  \bibinfo{journal}{Rev. Mod. Phys.} \textbf{\bibinfo{volume}{74}},
  \bibinfo{pages}{145} (\bibinfo{year}{2002}).

\bibitem[{\citenamefont{Bayer et~al.}(2000)\citenamefont{Bayer, Stern,
  Hawrylak, Fafard, and Forchel}}]{bayer:00a}
\bibinfo{author}{\bibfnamefont{M.}~\bibnamefont{Bayer}},
  \bibinfo{author}{\bibfnamefont{O.}~\bibnamefont{Stern}},
  \bibinfo{author}{\bibfnamefont{P.}~\bibnamefont{Hawrylak}},
  \bibinfo{author}{\bibfnamefont{S.}~\bibnamefont{Fafard}}, \bibnamefont{and}
  \bibinfo{author}{\bibfnamefont{A.}~\bibnamefont{Forchel}},
  \bibinfo{journal}{Nature} \textbf{\bibinfo{volume}{405}},
  \bibinfo{pages}{923} (\bibinfo{year}{2000}).

\bibitem[{\citenamefont{G{\'e}rard and Gayral}(1999)}]{gerard:99}
\bibinfo{author}{\bibfnamefont{J.~M.} \bibnamefont{G{\'e}rard}}
  \bibnamefont{and} \bibinfo{author}{\bibfnamefont{B.}~\bibnamefont{Gayral}},
  \bibinfo{journal}{Journal of Lightwave Technology}
  \textbf{\bibinfo{volume}{17}}, \bibinfo{pages}{2089} (\bibinfo{year}{1999}).

\bibitem[{\citenamefont{Michler et~al.}(2000)\citenamefont{Michler, Kiraz,
  Becher, Schoenfeld, Petroff, Zhang, Hu, and Imamoglu}}]{michler:00}
\bibinfo{author}{\bibfnamefont{P.}~\bibnamefont{Michler}},
  \bibinfo{author}{\bibfnamefont{A.}~\bibnamefont{Kiraz}},
  \bibinfo{author}{\bibfnamefont{C.}~\bibnamefont{Becher}},
  \bibinfo{author}{\bibfnamefont{W.~V.} \bibnamefont{Schoenfeld}},
  \bibinfo{author}{\bibfnamefont{P.~M.} \bibnamefont{Petroff}},
  \bibinfo{author}{\bibfnamefont{L.}~\bibnamefont{Zhang}},
  \bibinfo{author}{\bibfnamefont{E.}~\bibnamefont{Hu}}, \bibnamefont{and}
  \bibinfo{author}{\bibfnamefont{A.}~\bibnamefont{Imamoglu}},
  \bibinfo{journal}{Science} \textbf{\bibinfo{volume}{290}},
  \bibinfo{pages}{2282} (\bibinfo{year}{2000}).

\bibitem[{\citenamefont{Bayer et~al.}(2002)\citenamefont{Bayer, Ortner, Stern,
  Kuther, Gorbunov, Forchel, Hawrylak, Fafard, Hinzer, Reinecke
  et~al.}}]{bayer:02}
\bibinfo{author}{\bibfnamefont{M.}~\bibnamefont{Bayer}},
  \bibinfo{author}{\bibfnamefont{G.}~\bibnamefont{Ortner}},
  \bibinfo{author}{\bibfnamefont{O.}~\bibnamefont{Stern}},
  \bibinfo{author}{\bibfnamefont{A.}~\bibnamefont{Kuther}},
  \bibinfo{author}{\bibfnamefont{A.~A.} \bibnamefont{Gorbunov}},
  \bibinfo{author}{\bibfnamefont{A.}~\bibnamefont{Forchel}},
  \bibinfo{author}{\bibfnamefont{P.}~\bibnamefont{Hawrylak}},
  \bibinfo{author}{\bibfnamefont{S.}~\bibnamefont{Fafard}},
  \bibinfo{author}{\bibfnamefont{K.}~\bibnamefont{Hinzer}},
  \bibinfo{author}{\bibfnamefont{T.~L.} \bibnamefont{Reinecke}},
  \bibnamefont{et~al.}, \bibinfo{journal}{Phys. Rev. B}
  \textbf{\bibinfo{volume}{65}}, \bibinfo{pages}{195315}
  (\bibinfo{year}{2002}).

\bibitem[{\citenamefont{Young et~al.}(2006)\citenamefont{Young, Stevenson,
  Atkinson, Cooper, Ritchie, and Shields}}]{young:06}
\bibinfo{author}{\bibfnamefont{R.~J.} \bibnamefont{Young}},
  \bibinfo{author}{\bibfnamefont{R.~M.} \bibnamefont{Stevenson}},
  \bibinfo{author}{\bibfnamefont{P.}~\bibnamefont{Atkinson}},
  \bibinfo{author}{\bibfnamefont{K.}~\bibnamefont{Cooper}},
  \bibinfo{author}{\bibfnamefont{D.~A.} \bibnamefont{Ritchie}},
  \bibnamefont{and} \bibinfo{author}{\bibfnamefont{A.~J.}
  \bibnamefont{Shields}}, \bibinfo{journal}{New J. Phys.}
  \textbf{\bibinfo{volume}{8}}, \bibinfo{pages}{29} (\bibinfo{year}{2006}).

\bibitem[{\citenamefont{Hudson et~al.}(2007)\citenamefont{Hudson, Stevenson,
  Bennett, Young, Nicoll, Atkinson, Cooper, Ritchie, and Shields}}]{hudson:07}
\bibinfo{author}{\bibfnamefont{A.~J.} \bibnamefont{Hudson}},
  \bibinfo{author}{\bibfnamefont{R.~M.} \bibnamefont{Stevenson}},
  \bibinfo{author}{\bibfnamefont{A.~J.} \bibnamefont{Bennett}},
  \bibinfo{author}{\bibfnamefont{R.~J.} \bibnamefont{Young}},
  \bibinfo{author}{\bibfnamefont{C.~A.} \bibnamefont{Nicoll}},
  \bibinfo{author}{\bibfnamefont{P.}~\bibnamefont{Atkinson}},
  \bibinfo{author}{\bibfnamefont{K.}~\bibnamefont{Cooper}},
  \bibinfo{author}{\bibfnamefont{D.~A.} \bibnamefont{Ritchie}},
  \bibnamefont{and} \bibinfo{author}{\bibfnamefont{A.~J.}
  \bibnamefont{Shields}}, \bibinfo{journal}{Phys. Rev. Lett.}
  \textbf{\bibinfo{volume}{99}}, \bibinfo{pages}{266802}
  (\bibinfo{year}{2007}).

\bibitem[{\citenamefont{Gerardot et~al.}(2007)\citenamefont{Gerardot, Seidl,
  Dalgarno, Warburton, Granados, Garcia, Karrai, Badolato, and
  Petroff}}]{gerardot:07}
\bibinfo{author}{\bibfnamefont{B.~D.} \bibnamefont{Gerardot}},
  \bibinfo{author}{\bibfnamefont{S.}~\bibnamefont{Seidl}},
  \bibinfo{author}{\bibfnamefont{P.~A.} \bibnamefont{Dalgarno}},
  \bibinfo{author}{\bibfnamefont{R.~J.} \bibnamefont{Warburton}},
  \bibinfo{author}{\bibfnamefont{D.}~\bibnamefont{Granados}},
  \bibinfo{author}{\bibfnamefont{J.~M.} \bibnamefont{Garcia}},
  \bibinfo{author}{\bibfnamefont{K.}~\bibnamefont{Karrai}},
  \bibinfo{author}{\bibfnamefont{A.}~\bibnamefont{Badolato}}, \bibnamefont{and}
  \bibinfo{author}{\bibfnamefont{P.~M.} \bibnamefont{Petroff}},
  \bibinfo{journal}{Appl. Phys. Lett.} \textbf{\bibinfo{volume}{90}},
  \bibinfo{pages}{041101} (\bibinfo{year}{2007}).

\bibitem[{\citenamefont{Stace et~al.}(2003)\citenamefont{Stace, Milburn, and
  Barnes}}]{stace:03}
\bibinfo{author}{\bibfnamefont{T.~M.} \bibnamefont{Stace}},
  \bibinfo{author}{\bibfnamefont{G.~J.} \bibnamefont{Milburn}},
  \bibnamefont{and} \bibinfo{author}{\bibfnamefont{C.~H.~W.}
  \bibnamefont{Barnes}}, \bibinfo{journal}{Phys. Rev. B}
  \textbf{\bibinfo{volume}{67}}, \bibinfo{pages}{085317}
  (\bibinfo{year}{2003}).

\bibitem[{\citenamefont{Reimer et~al.}()\citenamefont{Reimer, Korkusinski,
  Lefebvre, Lapointe, Poole, Aers, Dalacu, M{cKinnon}, Frederick, Hawrylak
  et~al.}}]{reimer:07}
\bibinfo{author}{\bibfnamefont{M.~E.} \bibnamefont{Reimer}},
  \bibinfo{author}{\bibfnamefont{M.}~\bibnamefont{Korkusinski}},
  \bibinfo{author}{\bibfnamefont{J.}~\bibnamefont{Lefebvre}},
  \bibinfo{author}{\bibfnamefont{J.}~\bibnamefont{Lapointe}},
  \bibinfo{author}{\bibfnamefont{P.~J.} \bibnamefont{Poole}},
  \bibinfo{author}{\bibfnamefont{G.~C.} \bibnamefont{Aers}},
  \bibinfo{author}{\bibfnamefont{D.}~\bibnamefont{Dalacu}},
  \bibinfo{author}{\bibfnamefont{W.~R.} \bibnamefont{M{cKinnon}}},
  \bibinfo{author}{\bibfnamefont{S.}~\bibnamefont{Frederick}},
  \bibinfo{author}{\bibfnamefont{P.}~\bibnamefont{Hawrylak}},
  \bibnamefont{et~al.}, \emph{\bibinfo{title}{Voltage induced hidden symmetry
  and photon entanglement generation in a single site-selected quantum dot}},
  \bibinfo{howpublished}{arXiv:0706.1075}.

\bibitem[{\citenamefont{Avron et~al.}(2008)\citenamefont{Avron, Bisker,
  Gershoni, Lindner, Meirom, and Warburton}}]{avron:08}
\bibinfo{author}{\bibfnamefont{J.~E.} \bibnamefont{Avron}},
  \bibinfo{author}{\bibfnamefont{G.}~\bibnamefont{Bisker}},
  \bibinfo{author}{\bibfnamefont{D.}~\bibnamefont{Gershoni}},
  \bibinfo{author}{\bibfnamefont{N.~H.} \bibnamefont{Lindner}},
  \bibinfo{author}{\bibfnamefont{E.~A.} \bibnamefont{Meirom}},
  \bibnamefont{and} \bibinfo{author}{\bibfnamefont{R.~J.}
  \bibnamefont{Warburton}}, \bibinfo{journal}{Phys. Rev. Lett.}
  \textbf{\bibinfo{volume}{100}}, \bibinfo{pages}{120501}
  (\bibinfo{year}{2008}).

\bibitem[{\citenamefont{Santori et~al.}(2002)\citenamefont{Santori, Fattal,
  Vuckovic, Solomon, and Yamamoto}}]{santori:02b}
\bibinfo{author}{\bibfnamefont{C.}~\bibnamefont{Santori}},
  \bibinfo{author}{\bibfnamefont{D.}~\bibnamefont{Fattal}},
  \bibinfo{author}{\bibfnamefont{J.}~\bibnamefont{Vuckovic}},
  \bibinfo{author}{\bibfnamefont{G.~S.} \bibnamefont{Solomon}},
  \bibnamefont{and} \bibinfo{author}{\bibfnamefont{Y.}~\bibnamefont{Yamamoto}},
  \bibinfo{journal}{Nature} \textbf{\bibinfo{volume}{419}},
  \bibinfo{pages}{594} (\bibinfo{year}{2002}).

\bibitem[{\citenamefont{Santori et~al.}(2004)\citenamefont{Santori, Fattal,
  Vuckovic, Solomon, and Yamamoto}}]{santori:04}
\bibinfo{author}{\bibfnamefont{C.}~\bibnamefont{Santori}},
  \bibinfo{author}{\bibfnamefont{D.}~\bibnamefont{Fattal}},
  \bibinfo{author}{\bibfnamefont{J.}~\bibnamefont{Vuckovic}},
  \bibinfo{author}{\bibfnamefont{G.~S.} \bibnamefont{Solomon}},
  \bibnamefont{and} \bibinfo{author}{\bibfnamefont{Y.}~\bibnamefont{Yamamoto}},
  \bibinfo{journal}{Fortschr. Phys.} \textbf{\bibinfo{volume}{52}},
  \bibinfo{pages}{1180} (\bibinfo{year}{2004}).

\bibitem[{\citenamefont{Yuan et~al.}(2002)\citenamefont{Yuan, Kardynal,
  Stevenson, Shields, Lobo, Cooper, Beattie, Ritchie, and Pepper}}]{yuan:02}
\bibinfo{author}{\bibfnamefont{Z.}~\bibnamefont{Yuan}},
  \bibinfo{author}{\bibfnamefont{B.~E.} \bibnamefont{Kardynal}},
  \bibinfo{author}{\bibfnamefont{R.~M.} \bibnamefont{Stevenson}},
  \bibinfo{author}{\bibfnamefont{A.~J.} \bibnamefont{Shields}},
  \bibinfo{author}{\bibfnamefont{C.~J.} \bibnamefont{Lobo}},
  \bibinfo{author}{\bibfnamefont{K.}~\bibnamefont{Cooper}},
  \bibinfo{author}{\bibfnamefont{N.~S.} \bibnamefont{Beattie}},
  \bibinfo{author}{\bibfnamefont{D.~A.} \bibnamefont{Ritchie}},
  \bibnamefont{and} \bibinfo{author}{\bibfnamefont{M.}~\bibnamefont{Pepper}},
  \bibinfo{journal}{Science} \textbf{\bibinfo{volume}{295}},
  \bibinfo{pages}{102} (\bibinfo{year}{2002}).

\bibitem[{\citenamefont{Mandel and Wolf}(1995)}]{mandel:95}
\bibinfo{author}{\bibfnamefont{L.}~\bibnamefont{Mandel}} \bibnamefont{and}
  \bibinfo{author}{\bibfnamefont{E.}~\bibnamefont{Wolf}},
  \emph{\bibinfo{title}{Optical Coherence and Quantum Optics}}
  (\bibinfo{publisher}{Cambridge University Press},
  \bibinfo{address}{Cambridge}, \bibinfo{year}{1995}).

\bibitem[{\citenamefont{Troiani et~al.}(2006)\citenamefont{Troiani, Perea, and
  Tejedor}}]{troiani:06b}
\bibinfo{author}{\bibfnamefont{F.}~\bibnamefont{Troiani}},
  \bibinfo{author}{\bibfnamefont{J.~I.} \bibnamefont{Perea}}, \bibnamefont{and}
  \bibinfo{author}{\bibfnamefont{C.}~\bibnamefont{Tejedor}},
  \bibinfo{journal}{Phys. Rev. B} \textbf{\bibinfo{volume}{74}},
  \bibinfo{pages}{235310} (\bibinfo{year}{2006}).

\bibitem[{\citenamefont{Wooters}(1998)}]{wooters:98}
\bibinfo{author}{\bibfnamefont{W.~K.} \bibnamefont{Wooters}},
  \bibinfo{journal}{Phys. Rev. Lett.} \textbf{\bibinfo{volume}{80}},
  \bibinfo{pages}{2245} (\bibinfo{year}{1998}).

\bibitem[{\citenamefont{Jackson}(1999)}]{jackson:99}
\bibinfo{author}{\bibfnamefont{J.~D.} \bibnamefont{Jackson}},
  \emph{\bibinfo{title}{Classical Electrodynamics}}
  (\bibinfo{publisher}{Wiley}, \bibinfo{address}{New York},
  \bibinfo{year}{1999}).

\bibitem[{\citenamefont{Borri et~al.}(2001)\citenamefont{Borri, Langbein,
  Schneider, Woggon, Sellin, Ouyang, and Bimberg}}]{borri:01}
\bibinfo{author}{\bibfnamefont{P.}~\bibnamefont{Borri}},
  \bibinfo{author}{\bibfnamefont{W.}~\bibnamefont{Langbein}},
  \bibinfo{author}{\bibfnamefont{S.}~\bibnamefont{Schneider}},
  \bibinfo{author}{\bibfnamefont{U.}~\bibnamefont{Woggon}},
  \bibinfo{author}{\bibfnamefont{R.~L.} \bibnamefont{Sellin}},
  \bibinfo{author}{\bibfnamefont{D.}~\bibnamefont{Ouyang}}, \bibnamefont{and}
  \bibinfo{author}{\bibfnamefont{D.}~\bibnamefont{Bimberg}},
  \bibinfo{journal}{Phys. Rev. Lett.} \textbf{\bibinfo{volume}{87}},
  \bibinfo{pages}{157401} (\bibinfo{year}{2001}).

\bibitem[{\citenamefont{Hohenester et~al.}(2007)\citenamefont{Hohenester,
  Pfanner, and Seliger}}]{hohenester.prl:07}
\bibinfo{author}{\bibfnamefont{U.}~\bibnamefont{Hohenester}},
  \bibinfo{author}{\bibfnamefont{G.}~\bibnamefont{Pfanner}}, \bibnamefont{and}
  \bibinfo{author}{\bibfnamefont{M.}~\bibnamefont{Seliger}},
  \bibinfo{journal}{Phys. Rev. Lett.} \textbf{\bibinfo{volume}{99}},
  \bibinfo{pages}{047402} (\bibinfo{year}{2007}).

\bibitem[{\citenamefont{Breuer and Petruccione}(2002)}]{breuer:02}
\bibinfo{author}{\bibfnamefont{H.-P.} \bibnamefont{Breuer}} \bibnamefont{and}
  \bibinfo{author}{\bibfnamefont{F.}~\bibnamefont{Petruccione}},
  \emph{\bibinfo{title}{Open Quantum Systems}} (\bibinfo{publisher}{Oxford
  Univ. Press}, \bibinfo{address}{New York}, \bibinfo{year}{2002}).

\bibitem[{\citenamefont{Krummheuer et~al.}(2002)\citenamefont{Krummheuer, Axt,
  and Kuhn}}]{krummheuer:02}
\bibinfo{author}{\bibfnamefont{B.}~\bibnamefont{Krummheuer}},
  \bibinfo{author}{\bibfnamefont{V.~M.} \bibnamefont{Axt}}, \bibnamefont{and}
  \bibinfo{author}{\bibfnamefont{T.}~\bibnamefont{Kuhn}},
  \bibinfo{journal}{Phys. Rev. B} \textbf{\bibinfo{volume}{65}},
  \bibinfo{pages}{195313} (\bibinfo{year}{2002}).

\bibitem[{\citenamefont{Hohenester}(2007)}]{hohenester.jpb:07}
\bibinfo{author}{\bibfnamefont{U.}~\bibnamefont{Hohenester}},
  \bibinfo{journal}{J. Phys. B} \textbf{\bibinfo{volume}{40}},
  \bibinfo{pages}{S315} (\bibinfo{year}{2007}).

\bibitem[{\citenamefont{Walls and Millburn}(1995)}]{walls:95}
\bibinfo{author}{\bibfnamefont{D.~F.} \bibnamefont{Walls}} \bibnamefont{and}
  \bibinfo{author}{\bibfnamefont{G.~J.} \bibnamefont{Millburn}},
  \emph{\bibinfo{title}{Quantum Optics}} (\bibinfo{publisher}{Springer},
  \bibinfo{address}{Berlin}, \bibinfo{year}{1995}).

\bibitem[{\citenamefont{Gardiner and Zoller}(2004)}]{gardiner:04}
\bibinfo{author}{\bibfnamefont{C.~W.} \bibnamefont{Gardiner}} \bibnamefont{and}
  \bibinfo{author}{\bibfnamefont{P.}~\bibnamefont{Zoller}},
  \emph{\bibinfo{title}{Quantum Noise}} (\bibinfo{publisher}{Springer},
  \bibinfo{address}{Berlin}, \bibinfo{year}{2004}).

\bibitem[{\citenamefont{Kn{\"o}ll et~al.}(1984)\citenamefont{Kn{\"o}ll, Weber,
  and Sch{\"a}fer}}]{Knoell}
\bibinfo{author}{\bibfnamefont{L.}~\bibnamefont{Kn{\"o}ll}},
  \bibinfo{author}{\bibfnamefont{G.}~\bibnamefont{Weber}}, \bibnamefont{and}
  \bibinfo{author}{\bibfnamefont{T.}~\bibnamefont{Sch{\"a}fer}},
  \bibinfo{journal}{J. Phys. B: At. Mol. Phys.} \textbf{\bibinfo{volume}{17}},
  \bibinfo{pages}{4861} (\bibinfo{year}{1984}).

\bibitem[{not()}]{note.akopian}
\bibinfo{note}{This value is in reasonable agreement with the results of
  Ref.~\onlinecite{akopian:06} who reported a concurrence of $0.36\pm 0.10$ for
  a finestructure splitting of 27 $\mu$eV and a filter width of about 25
  $\mu$eV. With these values and the expressions given in the table, we obtain
  a concurrence of 0.12 for $\gamma_d=0$ and 0.1 for $\gamma_d=5$ $\mu$eV. This
  too small value in comparison to the experiment might be due to our use of a
  Lorentzian filter instead of the more square-like filter used in experiments.
  Decreasing the filter width we find an increase of the concurrence whose
  value, however, is rather insensitive to $\gamma_d$ due to the small value of
  the concurrence, in accordance to the results shown in the inset of Fig.~3(b)
  [see the large, flat areas of the concurrence for $\delta=20$ $\mu$eV and
  larger values of $\delta\omega$].}

\bibitem[{\citenamefont{Hafenbrak et~al.}(2007)\citenamefont{Hafenbrak, Ulrich,
  Michler, Wang, Rastelli, and Schmidt}}]{hafenbrak:07}
\bibinfo{author}{\bibfnamefont{R.}~\bibnamefont{Hafenbrak}},
  \bibinfo{author}{\bibfnamefont{S.~M.} \bibnamefont{Ulrich}},
  \bibinfo{author}{\bibfnamefont{P.}~\bibnamefont{Michler}},
  \bibinfo{author}{\bibfnamefont{W.}~\bibnamefont{Wang}},
  \bibinfo{author}{\bibfnamefont{A.}~\bibnamefont{Rastelli}}, \bibnamefont{and}
  \bibinfo{author}{\bibfnamefont{O.~G.} \bibnamefont{Schmidt}},
  \bibinfo{journal}{New J. Phys.} \textbf{\bibinfo{volume}{9}},
  \bibinfo{pages}{315} (\bibinfo{year}{2007}).

\end{thebibliography}

\end{document}